**Jet Impingement Cooling with Multi-stage Ducted Electroaerodynamic Actuators**

Quinna Nguyen[a]
C. Luke Nelson[b]
Daniel S. Drew[c, *]

*Corresponding author
[a]University of Utah
50 S Central Campus Dr, #2110
Salt Lake City, UT 84112, USA
quinna.nguyen@utah.edu

[b]University of Utah
50 S Central Campus Dr, #2110
Salt Lake City, UT 84112, USA
charles.nelson@utah.edu

[c]University of Hawaiʻi at Mānoa
2540 Dole St, Holmes Hall 483
Honolulu, HI 96822, USA
ddrew@hawaii.edu

## Abstract

Modern high-performance mobile electronics impose extreme constraints on thermal management, and traditional cooling methods often fail to meet requirements for power density, form factor, and durability. Jet impingement cooling offers a compelling solution but is typically hindered by the need for bulky ancillary hardware. Here, we demonstrate that compact arrays of reduced-scale electroaerodynamic (EAD) plasma actuators, which are silent, solid-state devices with no moving parts, can be used for direct jet impingement cooling of electronics. The main contribution is the first rigorous experimental demonstration and system-level validation of multi-stage, ducted electroaerodynamic jet arrays as a compact, fan-replacement impingement cooling solution for mobile electronics. We characterize the performance of both single- and multi-stage ducted actuators, including thermographic analysis of heat transfer coefficients and spatial cooling profiles. We also quantify the relationship between actuator stage count and cooling efficiency, showing that increasing the number of ion acceleration stages enhances jet velocity and heat transfer performance at a reduced efficiency. The actuators are then assembled into an array and directly compared to a conventional fan with similar coverage area, showing competitive performance at a fraction of the volume, weight, and power. Finally, we integrate the array onto a commercial edge AI system and show that thermal regulation during extended inference workloads matches that of a stock fan, without any moving mechanical components or noise. These results confirm that multi-stage EAD jet arrays are not only viable but advantageous for thermal management in mobile and high- performance systems, paving the way toward silent and miniaturized solid-state cooling solutions.

## Nomenclature

| Symbol | Description | Units |
|---|---|---|
| $A$ | area | $m^2$ |
| $E$ | electric field strength | $V\ m^{-1}$ |
| $\dot{E}$ | energy rate | $J\ s^{-1}$ |
| $F$ | electroaerodynamic force | N |
| $\mathbb{F}$ | electroaerodynamic force density | $N\ m^{-3}$ |
| $h$ | heat transfer coefficient | $W\ m^{-2}\ K^{-1}$ |
| $J$ | current density | $A\ m^{-2}$ |
| $k_n$ | thermal conductivity | $W\ m^{-1}\ K^{-1}$ |
| $q$ | heat flux | $W\ m^{-2}$ |
| $T$ | temperature | K |
| $u$ | jet velocity | $m\ s^{-1}$ |
| $\mu$ | ion mobility | $m^2\ V^{-1}\ s^{-1}$ |
| $\rho$ | air density | $kg\ m^{-3}$ |
| $V$ | voltage | V |
| $I$ | current | A |
| $\overline{Nu}$ | average Nusselt number | - |
| $Ra$ | Rayleigh number | - |
| $Pr$ | Prandtl number | - |
| $\epsilon$ | emissivity | - |
| $\sigma$ | Stefan-Boltzmann constant | $W\ m^{-2}\ K^{-4}$ |
| $COP$ | coefficient of performance | - |
| $\dot{Q}$ | heat removal rate | W |
| $z$ | distance of exhaust from the heated surface | m |
| $d$ | device exhaust diameter | m |

### *Subscripts*

| | |
|---|---|
| $conv$ | convection |
| $r$ | radiation |
| $cond$ | conduction |
| $gen$ | Joule heating input |
| $st$ | stored |
| $s$ | surface |
| $m$ | mean surface |
| $\infty$ | ambient |
| $in$ | energy into control volume |
| $out$ | energy out of control volume |
| $nc$ | natural convection |
| $f$ | foil |
| $p$ | paint |
| $L$ | characteristic length |
| $eff$ | efficiency |

## 1 Introduction

Increasing transistor counts and greater performance demands on processors have led to mobile electronics with high power densities and significant thermal management challenges [1], [2], [3]. Going forward, these challenges will be exacerbated by the continued growth of edge AI, robotics, and portable computing, where compact form factors, physical durability, and energy efficiency are all paramount. Excessive heat degrades performance and shortens the operational life of components, especially those relying on rare or costly materials (e.g., batteries, processors) [4]. For some emerging platforms with the most extreme demands on form factor, durability, and thermal stability, like AR and VR head-mounted displays (HMDs) and other wearable devices, thermal limitations are already a barrier to deployment.

Conventional electronics cooling systems, typically comprising rotary fans and metal heat sinks, cannot easily satisfy modern size-, weight-, and power- (SWaP) constraints [5]. At small scales, fans become aerodynamically inefficient and mechanically fragile, creating points of failure in systems designed for physical interaction or mobility. As a result, many devices now avoid active cooling entirely, relying on passive dissipation through design optimization [6]. While promising research exists in nanomaterials [7], liquid [8], [9], and thermoelectric [10] solutions, among others, their integration complexity often restricts use to larger or stationary platforms.

Jet impingement cooling offers a powerful and scalable alternative. By directing high-velocity jets toward hot surfaces, it generates strong local turbulence and thin thermal boundary layers, enabling enhanced heat transfer compared to conventional forced convection [11]. Most current implementations, however, require external pumping systems or pressurized fluid reservoirs, limiting applicability in compact electronics. We propose electrohydrodynamic (EHD) plasma actuators, specifically electroaerodynamic (EAD) devices which operate entirely on air, as a potential solution. In EAD devices, the bulk fluid flow is produced via the momentum-transferring collisions between accelerated ions and neutral air molecules. They therefore offer a silent, solid-state, and scalable pumping mechanism.

Electroaerodynamic actuators have been explored for fluid pumping in general, and for electronics cooling in particular, for several decades [12]. However, early implementations were limited by low flow rates, inefficient geometries, and the difficulty of integrating bulky or exposed electrodes into compact form factors. Recent advances in microfabrication and multi-stage ducted designs have enabled a new class of high-performance actuators capable of producing collimated jets with significantly higher momentum flux than previously documented devices [13], [14]. Multi-stage ducted (MSD) EAD thrusters increase thrust density by stacking ion acceleration regions in series, overcoming many of the limitations of earlier single-stage, open-air configurations [15], [16]. While these devices have shown promise for propulsion applications, their application to localized, high-heat-flux electronic cooling remains unexplored.

The primary novelty of this work is an experimental demonstration of the use of miniaturized multi-stage ducted electroaerodynamic actuators for direct jet impingement cooling of high-performance electronics. Unlike prior EAD cooling approaches that generated bulk airflow through conventional heat sink architectures, the present approach employs compact actuator arrays designed to deliver localized, high-momentum jets directly to heated surfaces. Individual actuators achieve peak heat transfer coefficients exceeding 300 W $m^{-2}$ $K^{-1}$ at millimeter-scale form factors, comparable to conventional fan-based cooling despite operating at volumetric flow rates more than an order of magnitude lower.

Key contributions include experimental thermographic characterization of spatial heat transfer distributions for multi-stage electroaerodynamic jets, quantitative evaluation of performance scaling with stage count and associated efficiency tradeoffs, direct comparison with conventional rotary fan cooling at equivalent power input, and system-level validation through integration with a commercial edge computing platform. These results establish multi-stage ducted electroaerodynamic actuators as a distinct and fundamentally more scalable architecture for compact, localized electronic cooling compared to prior electrohydrodynamic cooling approaches.

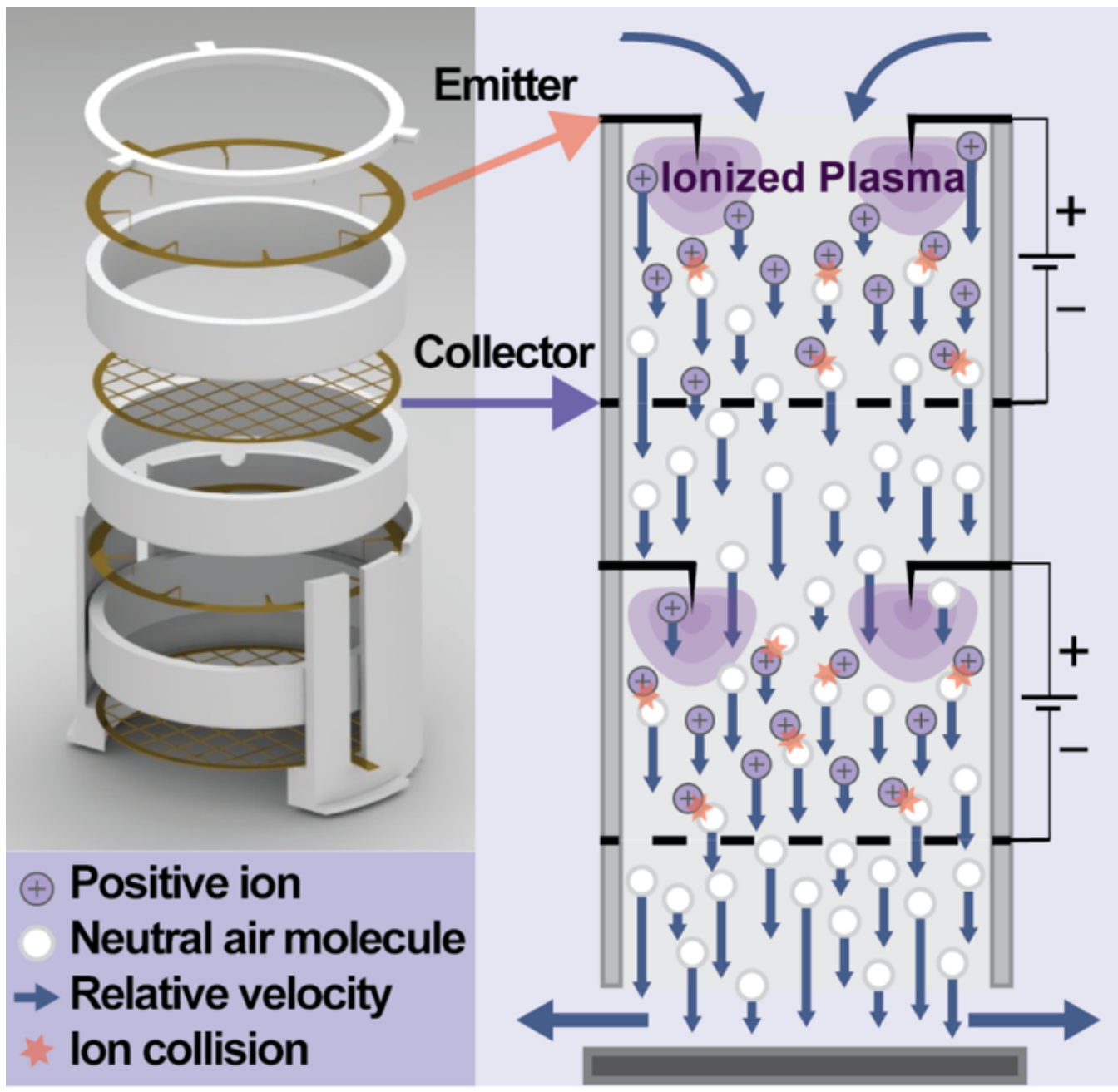

**Figure 1**. Render and operation schematic of a multi-stage ducted electroaerodynamic (EAD) actuator. Successive collisions with ions ejected from the corona plasma accelerate neutral air molecules throughout the ducted device. The collimated air impinges on the surface to be cooled once outside the duct and creates a wall jet, which carries away heat.

## 2 Technical Background

### 2.1 Corona discharge and electroaerodynamic force

Corona discharge refers to a self-sustaining atmospheric plasma generated at the surface of an “emitter” electrode when a high voltage is applied relative to a nearby “collector”. The resulting ionization relies on geometric asymmetry to prevent immediate dielectric breakdown of the intermediate air (Fig. 1). The discharge can be either positive or negative depending on the polarity of the emitter; in this work, we focus on positive corona discharges, which are associated with more uniform spatial distribution, lower ozone production, and smaller ionization regions at a given current [17]. In our devices, the corona plasma generates the ions which are subsequently accelerated into the drift region.

An electroaerodynamic (EAD) force arises from momentum transfer between drifting ions and neutral air molecules. Ions generated in the corona region are accelerated by the applied electric field and collide with neutrals, transferring momentum and creating a net fluid flow. A one-dimensional derivation for the total force acting on a volume of ions relates the ion travel distance, *d*, the ion current, *I*, and the ion mobility, *μ*. If momentum transferring collisions are frequent (valid at atmospheric pressure), the Coulomb force on the ions is approximately equal to the force imparted to the neutral air. As corona discharge-based EAD is a space charge limited effect governed by the Mott-Gurney (i.e., modified Child's Law) limit, the force can then be directly related to both the drift field magnitude *E* and the ion current *I* [18]:

$$F = qE = \frac{Id}{\mu} = \frac{9}{8}\epsilon_0 A E^2 \qquad (1)$$

Equation 1 indicates that the areal force density depends primarily on the applied electric field. One reason that continued miniaturization of EAD actuators to the microplasma regime is of interest is that their volumetric force density, F, scales inversely with the inter-electrode gap spacing:

$$\mathbb{F} = \frac{F_0}{A\,d} = \frac{9}{8}\epsilon_0 \frac{E^2}{d} \qquad (2)$$

Multi-stage ducted actuators extend this principle by introducing multiple ion acceleration regions in series. In each stage, ions generated at the emitter are accelerated toward a downstream collector, transferring momentum to the neutral air. Subsequent stages introduce additional electric field regions that continue to accelerate newly generated ions and impart further momentum to the already-moving airflow. This effectively increases the total impulse delivered to the fluid compared to a single-stage configuration operating at the same electric field strength. In addition to increasing cumulative momentum transfer, the duct structure confines and collimates the flow, reducing radial spreading and entrainment losses that would otherwise dissipate jet momentum in open-air geometries.

While Eqs. 1 and 2 are derived under a one-dimensional, steady, space-charge-limited approximation, this framework provides a useful first-order description of force generation and scaling in electroaerodynamic actuators. The derivation assumes uniform electric field distribution, negligible transverse gradients, and complete momentum transfer from ions to neutral molecules, conditions that are approximately satisfied within each individual acceleration gap when the ion mean free path is much smaller than the electrode spacing at atmospheric pressure. Under these conditions, the total body force imparted to the fluid depends primarily on ion current and electric field strength, and the Mott–Gurney limit provides an upper bound on achievable force density.

However, multi-stage ducted actuators introduce several additional physical effects not captured by this one-dimensional model. These include nonuniform electric field distributions near electrode edges, redistribution of space charge due to inter-stage flow coupling, aerodynamic drag on intermediate collector electrodes, viscous losses within the duct, and interaction between adjacent ion acceleration regions. These mechanisms reduce the effective momentum transfer efficiency per stage and lead to sub-ideal scaling of exhaust velocity and energy efficiency with stage count. Prior experimental and numerical studies have shown that, while total thrust and jet velocity can scale approximately linearly with the number of stages under optimized conditions, the electrical-to-mechanical conversion efficiency typically decreases due to these cumulative losses [13–16].

Accordingly, the one-dimensional Mott–Gurney formulation is employed here as a scaling framework to interpret trends in force density and actuator performance, rather than as a complete predictive model of flow and heat transfer. The experimentally measured heat transfer coefficients presented in Section 4 provide direct validation of actuator performance and inherently capture the aerodynamic and electrohydrodynamic effects not represented in the simplified analytical model.

### 2.2 Jet impingement cooling for electronics

Jet impingement cooling is a convective heat transfer technique where high-velocity fluid jets are directed perpendicularly at a target surface [19]. Upon impact, the fluid spreads into a wall jet, enhancing convective transport by disrupting thermal boundary layers. The method offers high local heat transfer coefficients, especially when the jet is confined or focused on hot spots. It is a promising alternative to traditional approaches to cooling for computing devices [20], [21]. Synthetic microjets have been shown to achieve thermal performance comparable to conventional electronic cooling fans for thermal hot-spots [22].

Most jet impingement implementations require active pumping, sealed ducts, or fluid recirculation systems, which are difficult to integrate in mobile electronics. Kalman et al. [23] and Owsenek et al. [24] previously proposed using corona discharge to generate jet-like flows as a solution. Their designs, however, lacked the scalability, efficiency, and performance needed for modern electronics.

### 2.3 Electrohydrodynamic devices for thermal management

Electrohydrodynamic force-induced flow has been investigated for use in thermal management since as early as 1963 [12]. Several reviews on the applications of EHD exist and cover the wide range of specific implementations and end-use cases which have been explored [12], [25], [26]. Here, we will note only a few of the most salient examples, focusing on convective cooling of electronics with relatively small EHD devices.

EAD cooling systems have historically been explored primarily as direct replacements for rotary fans by generating bulk airflow across heat sinks or internal electronic components. For example, Jewell-Larsen et al. demonstrated a functional laptop cooled by an electrohydrodynamic blower integrated within the same cavity as the stock rotary fan [27]. While this study established the feasibility of EAD-based cooling, the actuator operated as a distributed airflow generator rather than a localized jet source and therefore relied on conventional heat sink architectures for thermal transport. As a result, achievable heat transfer performance remained constrained by the relatively low momentum flux density and diffuse flow characteristics inherent to single-stage open-air EAD geometries. Furthermore, the physical dimensions of such blowers limited their applicability to ultra-compact or highly integrated electronics. Modern electronics also have highly asymmetric heating profiles, with small local hotspots which are inefficiently cooled by bulk convective flow solutions [2].

More recent efforts have investigated miniaturized EHD devices intended for localized cooling applications [28–31]. These studies demonstrated promising trends in heat transfer enhancement, but most were limited to either small heat loads, numerical predictions, or isolated experimental configurations without system-level integration. Prior devices also employed single-stage electrode configurations, which fundamentally limits achievable exhaust velocity and force density. Since electroaerodynamic force scales with the number of ion acceleration regions and inversely with electrode spacing (Section 2.1), multi-stage architectures provide a mechanism for increasing momentum transfer and improving cooling effectiveness. However, the use of multi-stage ducted electroaerodynamic actuators for direct jet impingement cooling of modern high-performance electronics has not been experimentally demonstrated until now.

## 3 Experimental Methods

### 3.1 Device design

The EAD devices used in this work are based on an annular multi-point-to-grid structure. The fabrication and assembly process for the individual actuators is as described in prior work investigating them for robot propulsion [14]. Briefly, the active electrodes (i.e., emitter and collector) are UV-laser micromachined (DPSS Samurai) out of 25 micrometer thick brass stock, the duct and other supporting mechanical components are SLA printed (Form 3, Clear Resin), and a printed jig and stamp-and-die set is used to aid in assembly. Multiple stages are stacked in series and fully enclosed in a duct, such that air is successively accelerated to higher velocities. Unless otherwise stated, the devices used in this work are each 10 mm in diameter, with 8 emitter tips positioned 2 mm from a collector grid, made of 25 μm thick brass, which has a 50 μm wire width and a 1 mm wire pitch. While prior work shows that mass flow output and efficiency, and therefore effectiveness as a cooling device, are strong functions of device geometry (e.g., collector grid spacing [32]), further investigation is not in the scope of this paper.

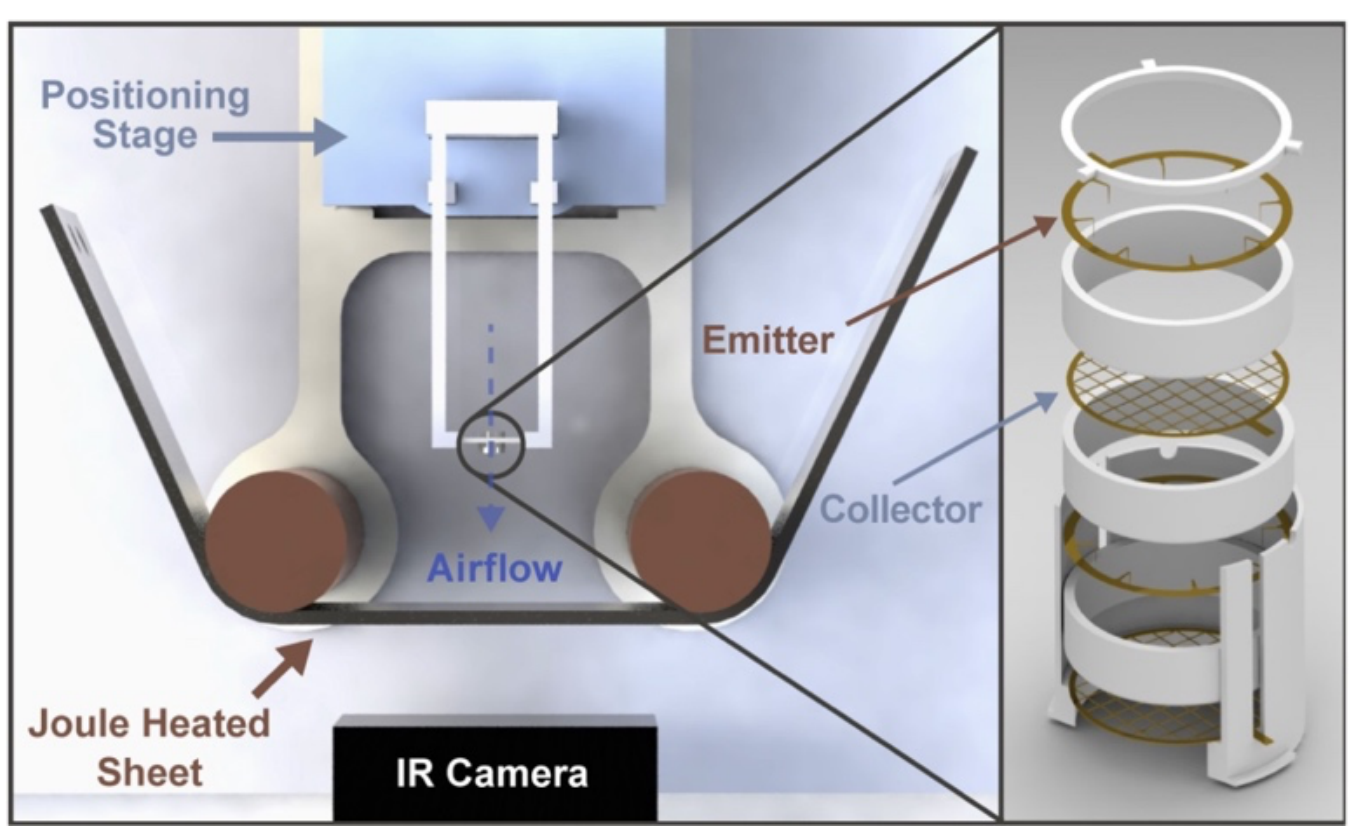

**Figure 2.** Render of the experimental test setup enabling direct thermographic analysis of cooling performance, with active electrodes of the EAD device labeled.

### 3.2 Experimental setup

A schematic of the experimental setup is shown in Figure 2. An infrared camera (PIR uc 605) is placed in front of a 50.8 micrometer stainless steel foil. The foil is held in tension against copper rod terminals by spring-loaded clips, pulled taut to decrease contact resistance and to prevent any deformation in the foil due to impinged air flow. On the camera-facing side of the sheet, a layer of paint (Zynolyte Hi-Temp Z635 matte black paint) is coated on the thin foil to provide a high emissivity surface of 0.94 [33]. A programmable power supply and programmable electronic load are placed in series with the foil to control Joule heating. Placed on the opposite side of the foil is an EAD cooling mount, with air blowing directly perpendicular to the sheet. At a given distance, the foil is held at a constant input power for Joule heating (via the programmable load) and applied voltage for EAD is increased as IR data is recorded.

### 3.3 Thermal analysis procedure

The analysis is based on the heated-thin-foil technique described by Stafford et al. [34], which uses the energy balance equation:

$$\dot{E}_{in} + \dot{E}_{gen} - \dot{E}_{out} = \dot{E}_{st} \quad (3)$$

This method isolates the forced convective heat transfer effects by accounting for secondary heat transfer losses such as radiation, natural convection, and conduction. To simplify the energy balance equation, experiments are run in steady state conditions such that $\dot{E}_{st}$ = 0. Based on research from Raghu et al. [35], we assume the same temperature field across the foil and paint thickness and layers due to a low Biot number. This assumption allows our analysis to be two-dimensional. Both foil and paint thicknesses were measured to account for conduction in the heat balance equation, which was shown to produce significant errors in the forced convection heat transfer coefficient calculations if ignored at this scale by Nogueira et al. [36].

The following equation defines the heat transfer coefficient as derived from the energy balance equation:

$$h_{fc} = \frac{q_{gen} - q_{nc} - q_r + q_c}{T_m - T_\infty} \qquad (4)$$

The foil is heated by the Joule effect, which can be used to calculate the heat flux as provided below:

$$q_{gen} = \frac{VI}{A} \qquad (5)$$

Edge effects from the natural convection and radiation produced at the copper rods, if not neglected, will result in non-uniform heat distribution during temperature gradient analysis. Thus, these effects are neglected and only the surface of the foil that dissipates heat by natural convection is measured experimentally in Eq. 6:

$$q_{nc} = \frac{\overline{Nu_L} k_{air} (T_m - T_\infty)}{L} \qquad (6)$$

The Nusselt number calculated in Eq. 6 utilizes the below formulation for laminar flow ($Ra \lesssim 10^9$) as measured on a vertical plate or foil [37]:

$$\overline{Nu_L} = 0.68 + \frac{0.670 {Ra_L}^{1/4}}{[1 + (0.492/Pr)^{9/16}]^{4/9}} \qquad (7)$$

Radiation heat flux is measured from the high emissivity paint layer with the following equation:

$$q_r = \varepsilon\sigma(T_m^4 - T_\infty^4) \qquad (8)$$

Findings from Stafford et al. [34] shows that neglecting paint effects can lead to significant errors when using the heated-thin-foil technique for lower heat transfer coefficient analysis. For this reason, calculation of tangential conduction measured along the length and height of the sheet includes both the foil and high-emissivity paint layer. We assume the thermal conductivity and thickness along both the foil and paint layer to be constant. Normal conduction along the thickness of the sheet is negligible. The equation for conduction is shown below:

$$q_c = (k_f t_f + k_p t_p)\left(\frac{\partial^2 T}{\partial x^2} - \frac{\partial^2 T}{\partial y^2}\right) \qquad (9)$$

Analyses near the edges of the heated sheet were considered unsuitable, as edge effects due to natural convection temperature gradients result in non-uniform heat transfer coefficients. The foil area is sized appropriately to safely leave out these edges and still capture the region of interest under active devices.

The relative contributions of parasitic heat transfer mechanisms in Eq. 4, including natural convection, radiation, and tangential conduction, were explicitly quantified for all device configurations. Across both single- and multi-stage actuators, forced convection associated with the electroaerodynamic jet accounted for approximately 85–90% of the total generated heat flux.

Natural convection and radiation contributed approximately 4–6% each, while tangential conduction provided a small correction of approximately −2 to −3%, with the negative sign indicating lateral heat redistribution within the foil rather than net heat loss. These proportions were consistent across device stage counts, confirming that the measured heat transfer coefficients are dominated by forced convection from the EAD jet.

The influence of parasitic heat transfer mechanisms was minimized through several experimental design choices, including the use of a thin stainless-steel foil to reduce tangential conduction, application of a high-emissivity coating to enable accurate radiative correction, and operation under steady-state conditions to eliminate transient energy storage effects. All secondary heat transfer terms were explicitly computed and subtracted from the total energy balance, ensuring that the reported heat transfer coefficients reflect only the forced convection contribution of the electroaerodynamic jet.

Uncertainty in the measured heat transfer coefficient was quantified using a propagation-of-uncertainty approach consistent with the methodology described by Stafford et al. [34]. Uncertainty contributions from each independent variable, including voltage (±5 mV), current (±5 mA), temperature measurement (±0.2 °C), emissivity (±0.02), geometric area (±1%), and natural convection correlation uncertainty (±15%), were evaluated by perturbing each parameter individually and recomputing the resulting heat transfer coefficient. The individual contributions were combined using a root-sum-square (RSS) method to obtain the overall uncertainty.

Across devices, the mean relative uncertainty in the heat transfer coefficient ranged from approximately 2–6%, depending on device stage count and local temperature gradients. The dominant sources of uncertainty were temperature measurement and geometric area estimation, while contributions from electrical input and radiation were comparatively small. These results confirm that the measured heat transfer coefficients are well resolved relative to measurement error and that the observed performance differences between stage counts exceed experimental uncertainty. These uncertainty measurements are omitted from the results in Section 4 for clarity.

## 4 Results and Discussion

### 4.1 Multi-stage ion jet coolers

We assessed the performance of individual devices (i.e., single jet actuators) and tested the hypothesis that multi-stage devices would exhibit higher heat transfer coefficients. The single-stage device design consists of a single multi-point emitter and collector electrode set, whereas the two-stage design has two layers of emitter-collector pairings (Fig. 2, right). We measured the exhaust velocity and current draw for single- and two-stage ion jet coolers (Fig. 3). As expected, exhaust velocity scales roughly with the square root of stage count, decreased from this ideal result by the effect of fluid drag. Here, the velocity was actually derived from force sensor measurements using simple momentum theory and the nominal duct exhaust area, but direct measurements have yielded similar results in prior work [14]. Future work using micro-PIV or MEMS-scale velocity probes may enable more precise direct characterization of the jet velocity field.

The peak exhaust velocity is approximately 3 m/s for the two-stage device at 3 kV, corresponding to a volumetric flow rate of approximately 0.5 CFM for a 10 mm diameter outlet. For comparison, the 40 mm reference fan used in this study (Section 4.2) has a manufacturer-rated maximum flow of 6.3 CFM. Despite this difference in volumetric flow rate, the measured peak heat transfer coefficients are comparable. These findings suggest that volumetric flow rate alone does not fully capture cooling effectiveness for localized jet-based cooling systems, which is consistent with the jet impingment cooling literature. Accordingly, this work focuses on directly measured heat transfer coefficients and heat removal rates as the primary performance metrics, which provide a more physically meaningful basis for comparison across different cooling modalities.

We measured the maximum heat transfer coefficient as a function of distance from the Joule-heated foil as well as spatially at a fixed distance (Fig. 4). Single two-stage devices achieve maximum heat transfer coefficients, at approximately the center of the duct exhaust, of around 350 $W/m^2K$. For reference, a standard computer fan has an average heat transfer coefficient between 25 and 250 $W/m^2K$ [38].

Thermal camera images for the one- and two-stage devices are shown in Figure 5. As applied voltage is increased past the corona onset voltage of approximately 2 kV, the jet plume is shown to qualitatively increase in size and strength, as expected from the results of Figure 3. The actual exhaust diameter of 10 mm is labeled on the images; as the highest heat transfer coefficient region extends far beyond this diameter, we assume that the high velocity impinged wall jet is efficiently carrying heat away before detaching from the surface. The foil is cooled down from a peak temperature over 100 degrees Celsius to a steady state value of under 50 degrees.

To quantitatively illustrate the multi-stage performance trade-off, we calculated both peak heat transfer performance and a power-normalized cooling metric for one- and two-stage devices at $z/d = 1.0$. Using the thermography-derived forced-convection heat flux $q_{fc}(x, y)$, we compute an area-integrated heat removal rate $\dot{Q}_{fc} = \sum q_{fc} A_{pix}$over a jet-footprint region of interest (duct diameter) and an expanded region of interest capturing the broader impingement cooling region. We then define a coefficient of performance for efficiency, $COP_{eff} = \dot{Q}_{fc}/P_{elec}$. The two-stage devices exhibit a higher near-peak averaged $h$ but a lower $COP_{eff}$ at the tested operating point of approximately 3 kV (Table 1). Bounding of the area-integrated heat removal rate implies, from the energy balance of Eq. 4, that both the one- and two-stage devices are sufficient to remove the generated Joule heating; exploring the limits of cooling capabilities remains as valuable future experimental work. An alternative (non-dimensionless) coefficient of performance $h/P_{elec}$ captures the improved performance of the two-stage device, while still showing the expected decrease with stage count.

The experimentally measured force values used to derive air velocity in Figure 3 were compared to one-dimensional analytical model predictions from Eq. 2 using measured current and nominal values for interelectrode gap $d = 2mm$ and ion mobility $\mu = 2 \times 10^{-4}\, m^2\, V^{-1} s^{-1}$. The single-stage actuator exhibited a deviation of approximately 11% from the theoretical prediction, indicating that the one-dimensional model provides a reasonable first-order estimate of force generation despite neglecting multidimensional aerodynamic and space charge effects. For multi-stage devices, an additional incremental loss of approximately 2.5% per stage was observed, consistent with prior experimental studies of ducted electroaerodynamic thrusters [14].

**Table 1.** Quantitative comparison table between one- and two-stage devices for both the isolated jet footprint (10 mm diameter) and expanded jet impingement region (roughly 30 mm diameter).

| n | **Voltage** (kV) | **Current** (μA) | **Electrical Power** (W) | **h** ($W/m^2\cdot K$) | $\dot{Q}_{fc}$**,jet** (W) | $\dot{Q}_{fc}$**,full** (W) | $COP_{eff}$ jet | $COP_{eff}$ full | $h/P_e$ ($m^{-2}\cdot K^{-1}$) |
|---|---|---|---|---|---|---|---|---|---|
| 1 | 3.01 | 37.9 | 0.114 | 251.4 | 0.46 | 3.76 | 4.03 | 32.95 | 2204 |
| 2 | 2.98 | 67.2 | 0.200 | 337.7 | 0.45 | 3.79 | 2.25 | 18.90 | 1686 |

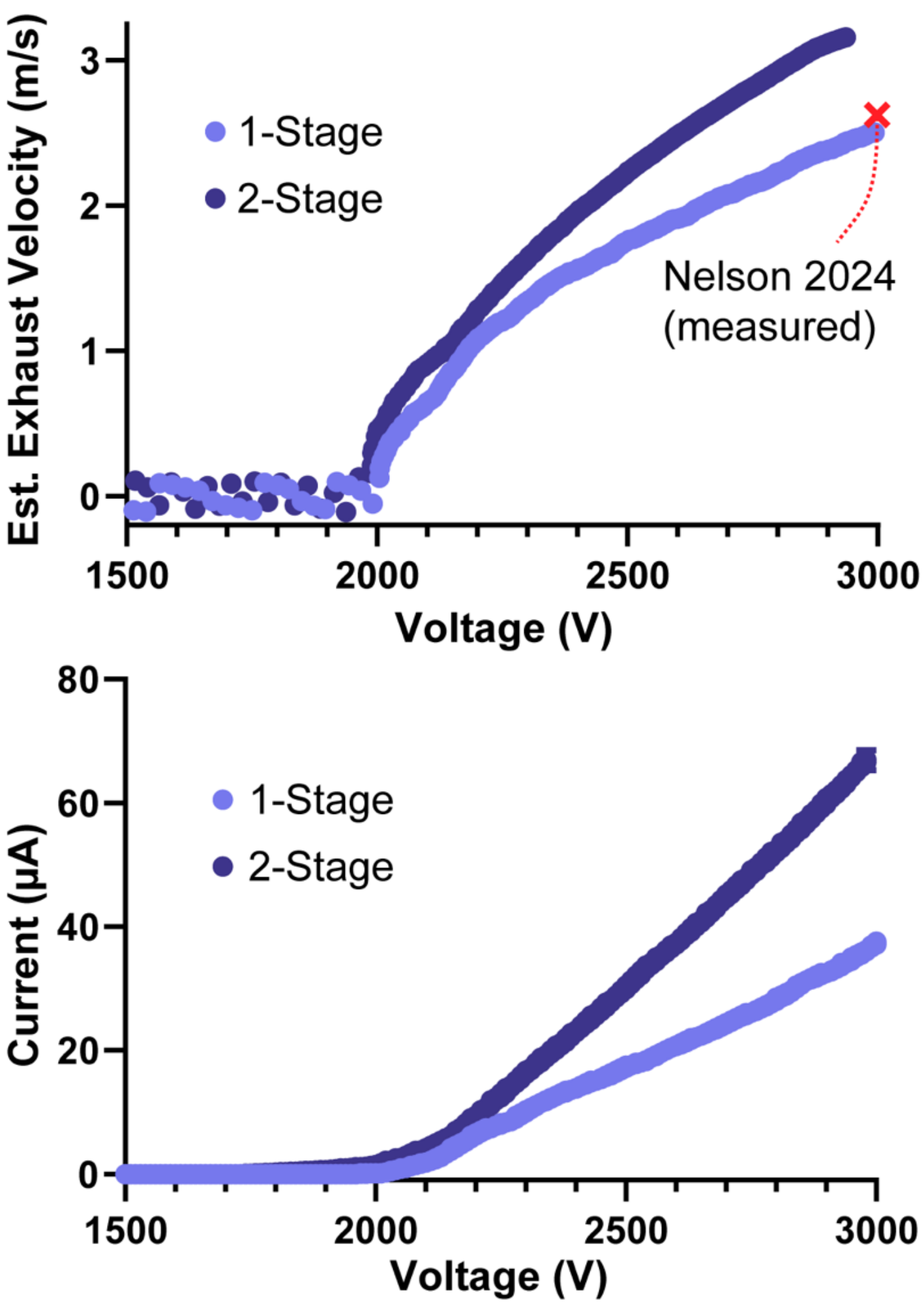

**Figure 3:** One and two-stage performance measurements. **Top:** Exhaust velocity, estimated from momentum theory and direct force measurements, versus applied voltage. Single measurement point from equivalent device in prior work confirms approximate validity of the velocity estimation. **Bottom:** Ion current versus applied voltage. Note that, although these devices have high operating voltages, the required currents are extremely small.

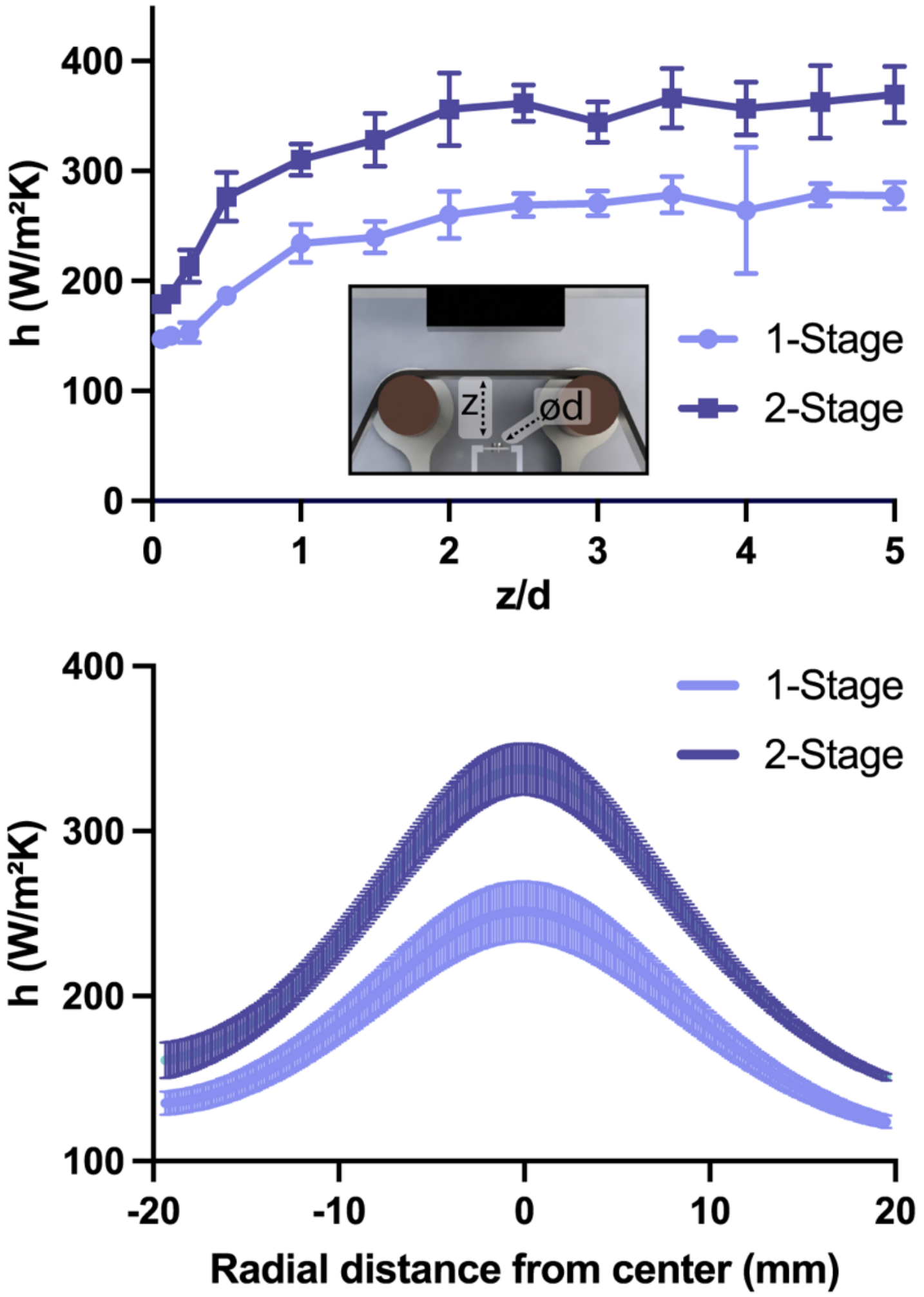

**Figure 4.** Heat transfer coefficient for one and two-stage devices. **Top:** Heat transfer coefficient as a function of the dimensionless *z/d* parameter, where *z* is the distance from the sheet and *d* is the device diameter (see inset). **Bottom:** Heat transfer coefficient as a function of radial distance from the jet exhaust center in one axis; operation of device was performed at 3 kV. The data was qualitatively axisymmetric. Error bars are standard deviation across three trials per device.

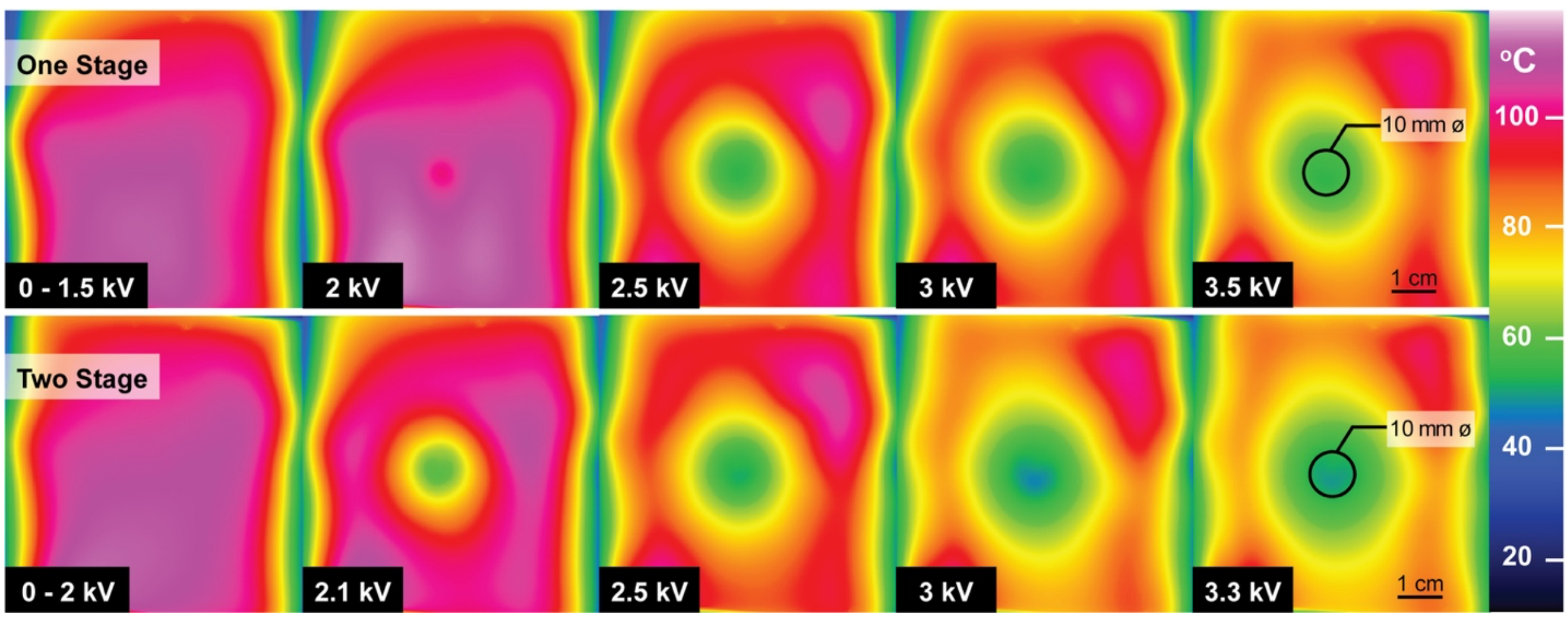

**Figure 5.** Thermal images showing temperature profile versus applied voltage for a one and two-stage device, with exhaust diameter labeled.

### 4.2 Jet array and comparison to conventional cooler

We created a functional prototype for drop-in replacement of a standard cooling fan by arraying four of the jet actuators, characterized individually in Section 4.1, and connecting them with a solid plate to channel the resultant convective flow. The stock fan used for comparison was a 40 mm x 40 mm x 20 mm PWM-controlled brushless DC axial fan (Waveshare FAN-4020-PWM-5V), with a rated voltage of 5 V, maximum rotational speed of 7500 RPM, maximum airflow of 6.3 CFM, and maximum power consumption of 1.6 W, which is sold as the default fan for the NVIDIA Jetson Nano. The “jet array” is comparable in size to the stock fan (Fig. 6, top) and produces a qualitatively similar thermal profile on the heated foil sheet (Fig. 6, bottom). Notably, the active area for the jet array does not exhibit the same “dead zone” induced by the fan hub. We measured the radial heat transfer coefficient and showed that, quantitatively, the jet array achieves a higher maximum heat transfer coefficient and has a much more uniform cooling profile (Fig. 7). The results shown in Figure 6 and Figure 7 are at an equivalent input power, corresponding to 0.45 W for the single-stage jet array and 0.8 W for the two-stage array. This is below the maximum output of the fan (approx. 30% and 50% duty cycle, respectively) but corresponds well to the nominal cooling PWM specified by the Jetson Nano (discussed in Section 4.3) integration guide; in practice, the fan is not actually run at max output for cooling this edge device. A video of the cooling process which resulted in Figure 6 can be seen in Supplementary Video 1.

The testing performed here was focused on relative cooling efficiency rather than maximum achievable cooling capacity. Figure 6 provides evidence that a two-stage jet array has a higher $h_{avg}/\,P_{elec}$ coefficient of performance at this specific operating point. When operated at maximum output, however, the fan would certainly have a higher average heat transfer coefficient than the jet array. To match that heat transfer coefficient, the number of stages could be increased, though this would introduce an efficiency penalty (see Section 4.1), increased volume, and other potential challenges (e.g., increased ozone generation). Exploring both the limits of this tradeoff and new device designs which change things like efficiency loss per stage remains as valuable future work, especially as edge compute devices continue to increase in power density.

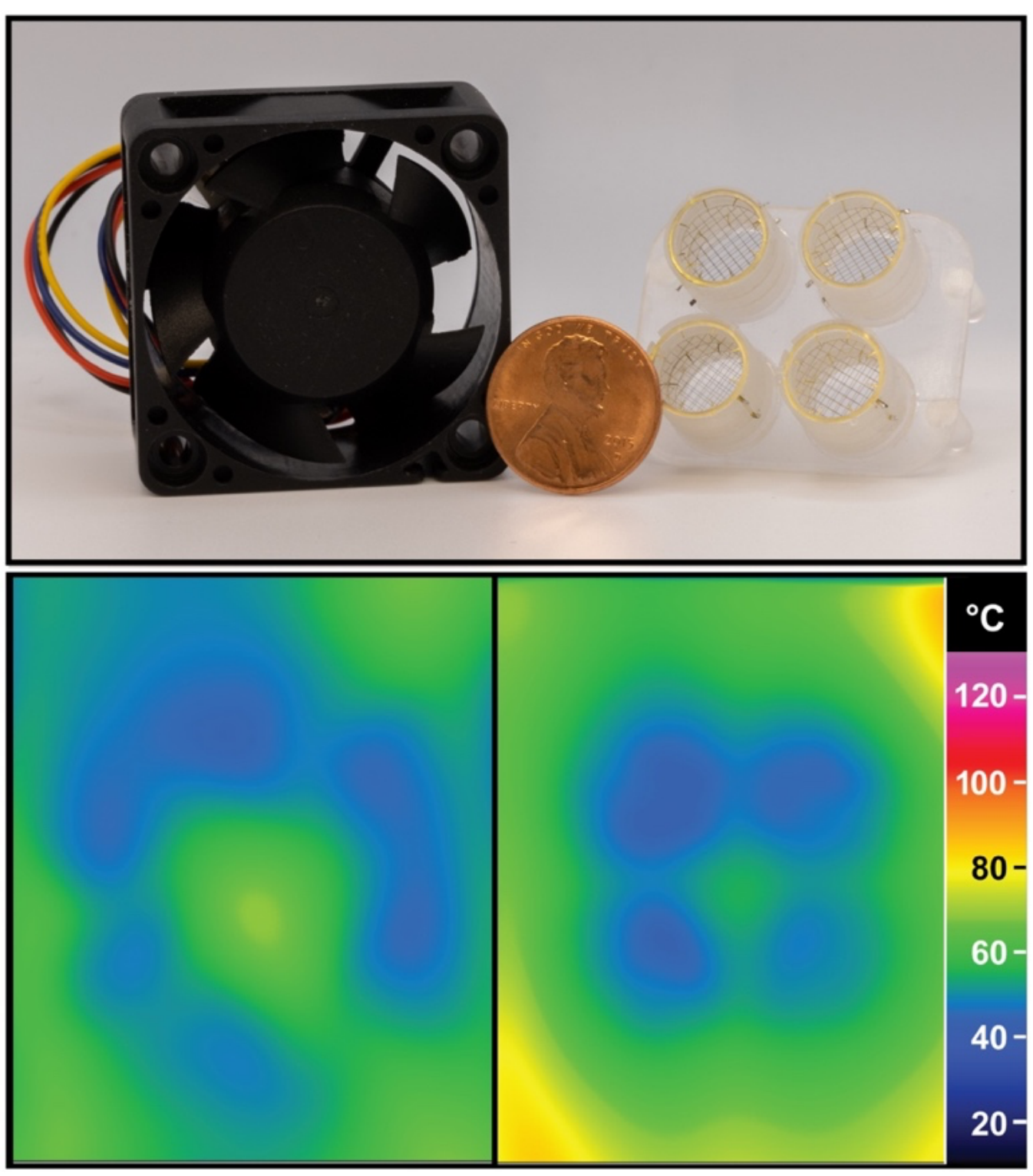

**Figure 6.** Direct comparison of devices. **Top:** Fan next to the quad-array. **Bottom:** Cooling of a heated sheet using the fan (left) and the quad-array (right) at an equivalent input power.

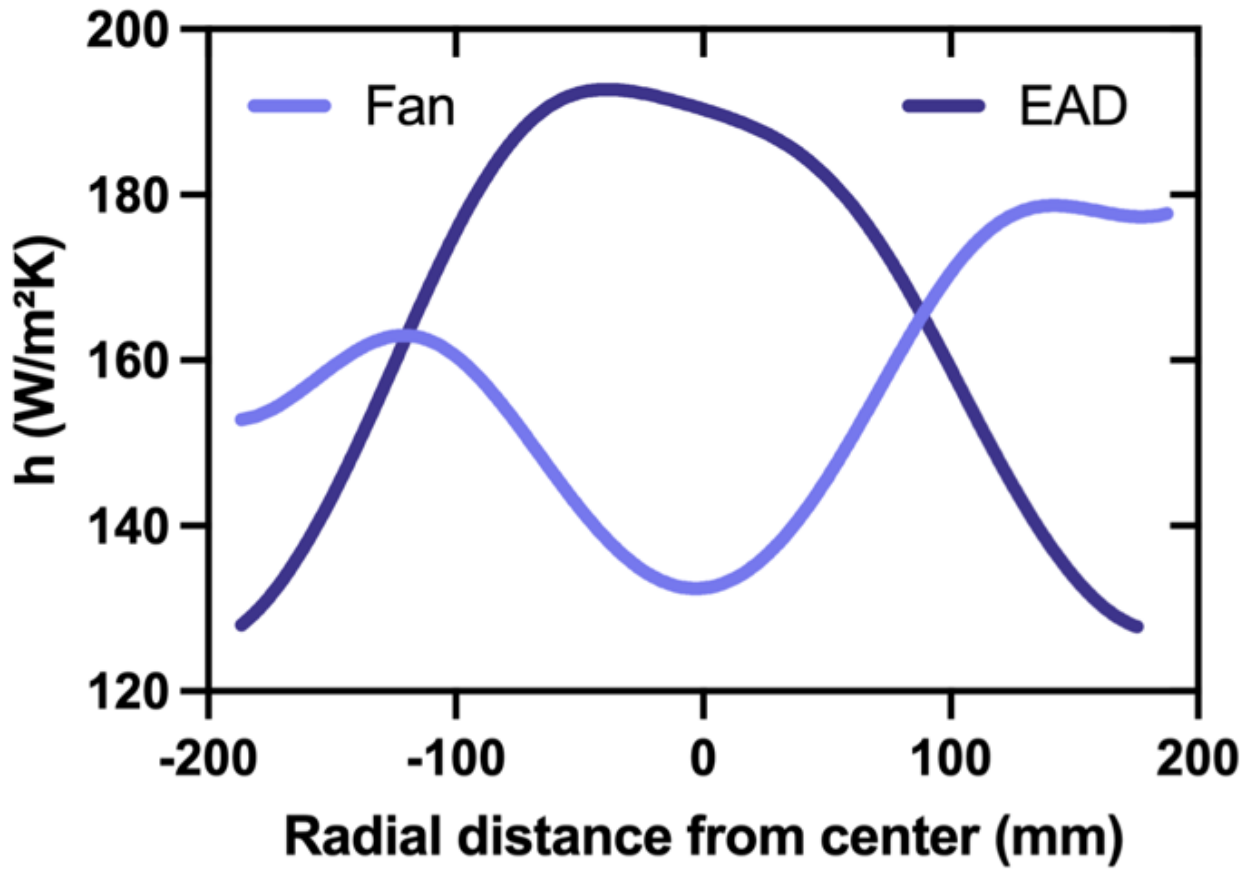

**Figure 7.** Heat transfer coefficient as a function of radial distance from the exhaust center for an EAD jet array and a conventional fan at equivalent input power.

**4.3 Integration and testing with high-performance mobile electronics**

To prove that the electroaerodynamic jet array is suitable for real electronics cooling applications, we integrated and tested it with an NVIDIA Jetson Nano edge AI computer, and compared the performance under actual loads as emulated with machine learning benchmarks. For example, without any thermal management, the Jetson quickly reaches unstable temperatures and can automatically shut off during standard computer vision benchmarks (Fig. 8, top). We designed the jet array such that the exhausts would be centered on the highest temperature regions and included mounting standoffs compatible with the existing spaces on the board (Fig. 8, middle). The array can then be screwed directly onto the Jetson using the same mounting hardware as the stock fan (Fig. 8, bottom).

Using the onboard sensor to measure the temperature allowed us to directly compare performance under load using no cooler, the stock fan, and our EAD jet array. We ran a suite of eight computer vision benchmarks in succession, with no time for the Jetson to cool between tests. The entire suite took over 100 minutes to complete. Figure 9 shows the temperature values versus time (top) and the average benchmark frames-per-second score (bottom). The EAD array performs nearly identically to the stock fan, as expected from the thermal characterization in the preceding sections. The jet array is silent, has no mechanical moving parts which could degrade over time or fail catastrophically from a fall or collision, and is 84% lighter (4.5 g, versus the 28.2 g fan). See Supplementary Video 2 for a demonstration of the silent and solid-state nature of operation. Notably, there is no observed degradation in performance over the course of the 100+ minute test suite.

While stable multi-hour operation provides initial evidence of robustness, assessing this reliability across longer durations and a wider range of environmental factors remains valuable future work. Humidity-induced electrode corrosion, for example, could alter electrode geometry and conductivity, leading to degradation of actuator performance [39]. Dust accumulation on the emitter or collector electrodes, with the latter especially susceptible due to the electrostatic precipitation effect of EAD devices, could decrease performance by reducing the effective electric field strength and partially obstructing airflow within the duct [40]. Ambient air pressure and humidity can also strongly affect performance in non-destructive ways by altering ion mobility, breakdown voltage, and momentum transfer efficiency [41], which has implications for cooling efficacy in different geographic regions and operating environments. The devices in this work were not subject to rapid degradation under typical laboratory conditions as has been documented for other micro-scale EAD electrodes [42], supporting their viability for further long-duration reliability studies.

An important consideration for future field testing with consumer electronics is safety. One of the reaction byproducts of the corona ionization process is $O_3$, ozone. Exposure to ozone is deleterious to human health in relatively low concentrations [43]. While not within the primary scope of this manuscript, there are several design factors which mitigate the issue of ozone generation from the devices described here. The millimeter-scale actuators operate at relatively low total discharge current and confine the ionization region within a ducted geometry, both of which are expected to limit overall ozone production compared to larger, open-air EHD blower systems. Additionally, positive corona discharge generates significantly less ozone than negative discharges [44], sharp/fine emitter electrodes have been shown experimentally to significantly reduce ozone production [45], and elevated emitter electrode temperatures have been shown to reduce ozone production by as much as 85% [46]. Directly measuring the ozone production of the multi-stage ducted actuators under representative operating conditions, and exploring mitigation strategies (e.g., inclusion of granular activated carbon [47]), remains valuable future work necessary to fully characterize their suitability for widespread deployment.

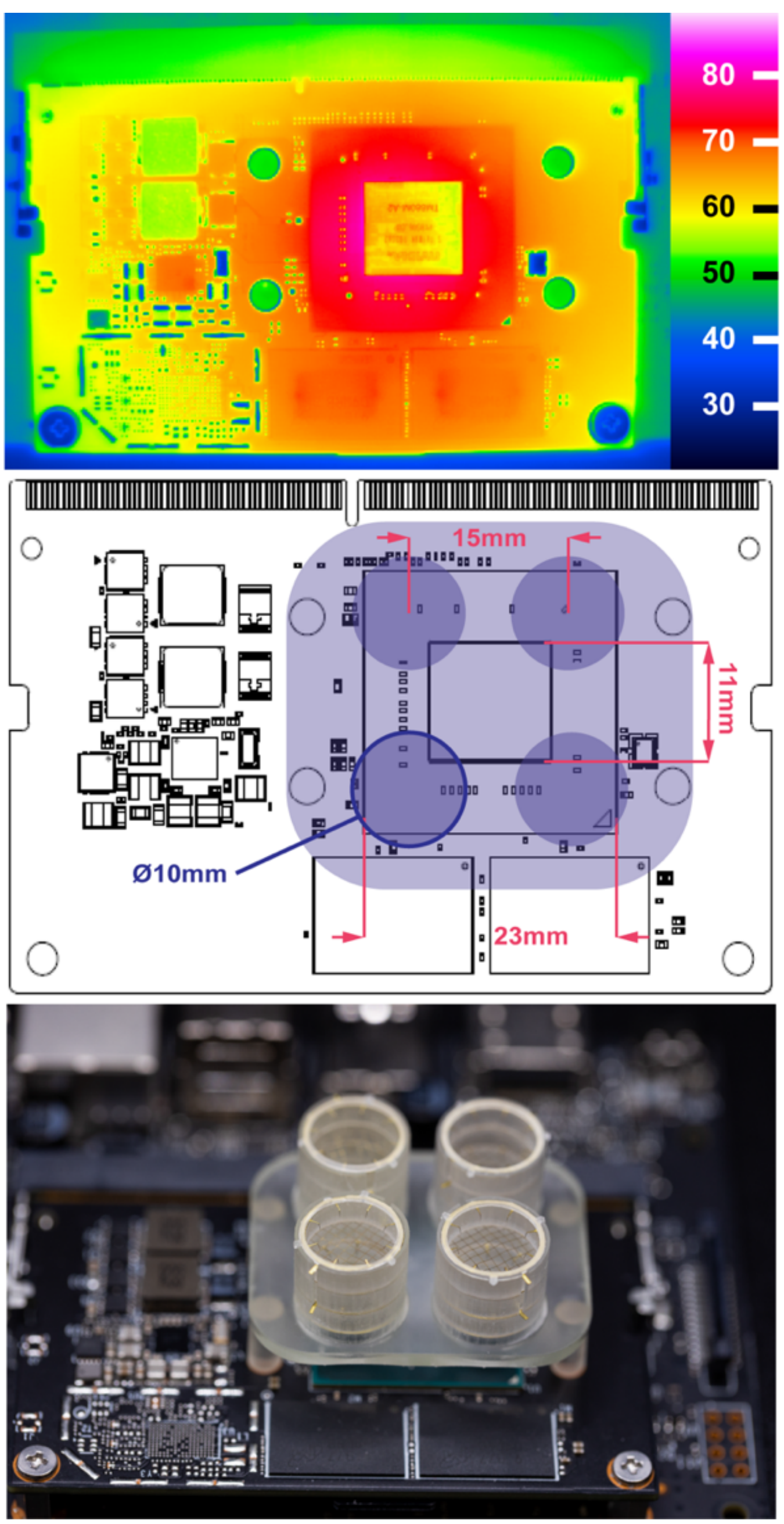

**Figure 8:** Hotspot analysis and direct integration of EAD quad-array with an edge computing system. **Top:** Thermal images show the hottest area of the Jetson Nano during a standard computer vision benchmark. **Middle:** Dimensioned drawing showing placement of the four EAD jet actuators and supporting plate on the Jetson. **Bottom:** Image of the integrated cooling device attached using the pre-existing mounting holes.

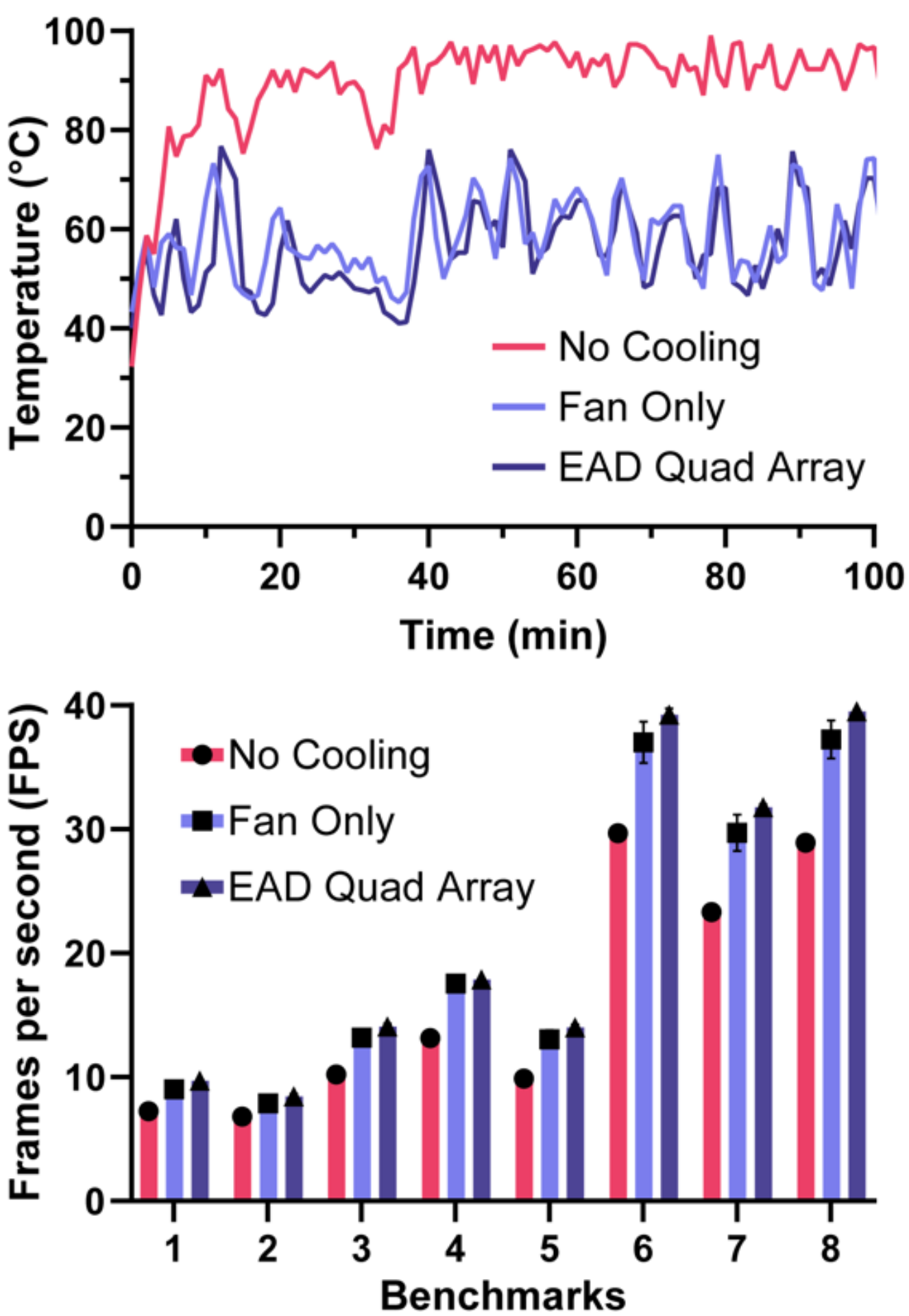

**Figure 9.** Comparing cooling performance under load. **Top**: Temperature data reported by the onboard sensor during completion of a computer vision benchmark. **Bottom:** Processed frames per second for different typical benchmarking tasks as a function of cooling strategy.

## 5 Conclusions

The efficacy of jet impingement cooling with multi-stage electroaerodynamic actuators was experimentally assessed. The specific geometric design of the millimeter-scale annular devices used in this work was shown to result in high cooling performance (i.e., a heat transfer coefficient of over 300 W/m$^2$K) at operating voltages of approximately 3 kV. When integrated directly onto the thermal hotspot of a modern edge computing device, an array of four EAD actuators proved capable of near-identical cooling as compared to a standard fan, while being significantly smaller, lighter weight, having no moving parts, and operating in silence. Thermography also shows that EAD jets have qualitatively more uniform flow and cooling distributions at their outlet as compared to fans (e.g., they have no cold spot beneath the rotor hub), indicating they are better suited for further miniaturization and direct integration onto small integrated circuits. Critically, this work experimentally demonstrates for the first time that compact multi-stage ducted electroaerodynamic actuators can provide direct jet impingement cooling of modern high-power electronics with performance comparable to conventional fan-based solutions, while enabling silent, solid-state operation in substantially reduced volume and mass. Further, our devices are able to operate for hours at a time without performance degradation. This work is an important step towards deeper integration (e.g., including the high voltage conversion circuitry on-board) and more rigorous testing (e.g., including environmental factors and motion) which would prove electroaerodynamic jet impingement cooling is a truly viable alternative for convection cooling of modern electronics.

## Glossary

**Biot number (Bi)** — Ratio of internal conductive to external convective heat resistance.
**CFM (Cubic Feet per Minute)** — Volumetric airflow rate measure.
**Collector** — Electrode that receives ions after acceleration.
**Corona discharge** — Plasma formed when high electric fields ionize air near an electrode.
**Drift region** — Space between electrodes where ions accelerate and drive airflow.
**Electroaerodynamic (EAD)** — Airflow generation by ion-neutral momentum transfer in air.
**Electrohydrodynamic (EHD)** — Electrically induced fluid motion in gases or liquids.
**Electrostatic Fluid Accelerator (EFA)** — Device converting electric energy to neutral gas motion.
**Emitter** — High-voltage electrode that initiates ionization.
**Heat transfer coefficient (h)** — Convective heat flux per unit area and temperature difference.
**High emissivity paint** — Coating with high IR emissivity for thermographic accuracy.
**Ion mobility (μ)** — Ion velocity per unit electric field strength.
**Jet array** — Assembly of multiple actuators producing distributed jets.
**Jet impingement cooling** — Surface cooling via direct high-velocity jet impact.
**Joule heating** — Heat generation by electric current through resistance.
**Multi-stage actuator** — EAD device with serial emitter–collector pairs for higher velocity.
**Nusselt number (Nu)** — Ratio of convective to conductive heat transfer.
**Positive corona discharge** — Corona mode with positive emitter bias relative to collector.
**Rayleigh number (Ra)** — Dimensionless measure of natural convection onset.
**SLA printing (Stereolithography)** — Resin 3D printing method for precise components.
**Stage count** — Number of acceleration stages in a multi-stage actuator.
**Thermography (Infrared)** — Imaging surface temperature using IR radiation.
**Wall jet** — Radial flow spreading along a surface after jet impact.

## Bibliography

[1] A.L. Moore, L. Shi, Emerging challenges and materials for thermal management of electronics, Mater. Today 17(4) (2014) 163–174. https://doi.org/10.1016/j.mattod.2014.04.003.

[2] Y. Hang, H. Kabban, Thermal management in mobile devices: challenges and solutions, in: Proc. 31st Thermal Measurement, Modeling and Management Symp. (SEMI-THERM), 2015, pp. 46–49. https://doi.org/10.1109/SEMI-THERM.2015.7100138.

[3] S.V. Garimella, et al., Thermal challenges in next-generation electronic systems, IEEE Trans. Compon. Packag. Technol. 31(4) (2008) 801–815. https://doi.org/10.1109/TCAPT.2008.2001197.

[4] V. Lakshminarayanan, N. Sriraam, The effect of temperature on the reliability of electronic components, in: Proc. IEEE Int. Conf. Electron., Comput. Commun. Technol. (CONECCT), 2014, pp. 1–6. https://doi.org/10.1109/CONECCT.2014.6740182.

[5] R. Grimes, E. Walsh, P. Walsh, Active cooling of a mobile phone handset, Appl. Therm. Eng. 30(16) (2010) 2363–2369. https://doi.org/10.1016/j.applthermaleng.2010.06.002.

[6] H. Liu, C. Yang, R. Wang, Passive thermal management of electronic devices, Device 3(3) (2025) 100684. https://doi.org/10.1016/j.device.2024.100684.

[7] D. Thesiya, H. Patel, G.S. Patange, A comprehensive review of electronic cooling: a nanomaterial perspective, Int. J. Thermofluids 19 (2023) 100382. https://doi.org/10.1016/j.ijft.2023.100382.

[8] A. Bar-Cohen, M. Arik, M. Ohadi, Direct liquid cooling of high flux micro and nano electronic components, Proc. IEEE 94(8) (2006) 1549–1570. https://doi.org/10.1109/JPROC.2006.879791.

[9] H.Y. Zhang, D. Pinjala, P.S. Teo, Thermal management of high power dissipation electronic packages: from air cooling to liquid cooling, in: Proc. 5th Electron. Packag. Technol. Conf. (EPTC), 2003, pp. 620–625. https://doi.org/10.1109/EPTC.2003.1271593.

[10] H. Kattan, S.W. Chung, J. Henkel, H. Amrouch, On-demand mobile CPU cooling with thin-film thermoelectric array, IEEE Micro 41(4) (2021) 67–73. https://doi.org/10.1109/MM.2021.3061335.

[11] M. Chaudhari, B. Puranik, A. Agrawal, Heat transfer characteristics of synthetic jet impingement cooling, Int. J. Heat Mass Transf. 53(5–6) (2010) 1057–1069. https://doi.org/10.1016/j.ijheatmasstransfer.2009.11.005.

[12] E.D. Fylladitakis, M.P. Theodoridis, A.X. Moronis, Review on the history, research, and applications of electrohydrodynamics, IEEE Trans. Plasma Sci. 42(2) (2014) 358–375. https://doi.org/10.1109/TPS.2013.2297173.

[13] D.S. Drew, S. Follmer, High force density multi-stage electrohydrodynamic jets using folded laser microfabricated electrodes, in: Proc. IEEE Int. Conf. Solid-State Sens., Actuators Microsyst. (Transducers), 2021, pp. 54–57. https://doi.org/10.1109/Transducers50396.2021.9495704.

[14] C.L. Nelson, D.S. Drew, High aspect ratio multi-stage ducted electroaerodynamic thrusters for micro air vehicle propulsion, IEEE Robot. Autom. Lett. 9(3) (2024) 2702–2709. https://doi.org/10.1109/LRA.2024.3355728.

[15] N. Gomez-Vega, A. Brown, H. Xu, S.R.H. Barrett, Model of multistaged ducted thrusters for high-thrust-density electroaerodynamic propulsion, AIAA J. 61(2) (2023) 767–779. https://doi.org/10.2514/1.J061948.

[16] N. Gomez-Vega, S.R.H. Barrett, Order-of-magnitude improvement in electroaerodynamic thrust density with multistaged ducted thrusters, AIAA J. (2024). https://doi.org/10.2514/1.J063431.

[17] J.S. Chang, P.A. Lawless, T. Yamamoto, Corona discharge processes, IEEE Trans. Plasma Sci. 19(6) (1991) 1152–1166. https://doi.org/10.1109/27.125038.

[18] L. Pekker, M. Young, Model of ideal electrohydrodynamic thruster, J. Propuls. Power 27(4) (2011) 786–792. https://doi.org/10.2514/1.B34097.

[19] B. Kwon, T. Foulkes, T. Yang, N. Miljkovic, W.P. King, Air jet impingement cooling of electronic devices using additively manufactured nozzles, IEEE Trans. Compon. Packag. Manuf. Technol. 10(2) (2020) 220–229. https://doi.org/10.1109/TCPMT.2019.2936852.

[20] R.C. Chu, R.E. Simons, M.J. Ellsworth, R.R. Schmidt, V. Cozzolino, Review of cooling technologies for computer products, IEEE Trans. Device Mater. Reliab. 4(4) (2004) 568–585. https://doi.org/10.1109/TDMR.2004.840855.

[21] S.M.S. Murshed, C.A. Nieto de Castro, A critical review of traditional and emerging techniques and fluids for electronics cooling, Renew. Sustain. Energy Rev. 78 (2017) 821–833. https://doi.org/10.1016/j.rser.2017.04.112.

[22] D.S. Kercher, J.B. Lee, O. Brand, M.G. Allen, A. Glezer, Microjet cooling devices for thermal management of electronics, IEEE Trans. Compon. Packag. Technol. 26(2) (2003) 359–366. https://doi.org/10.1109/TCAPT.2003.815116.

[23] H. Kalman, E. Sher, Enhancement of heat transfer by means of a corona wind created by a wire electrode and confined wings assembly, Appl. Therm. Eng. 21(3) (2001) 265–282. https://doi.org/10.1016/S1359-4311(00)00038-7.

[24] B.L. Owsenek, J. Seyed-Yagoobi, R.H. Page, Experimental investigation of corona wind heat transfer enhancement with a heated horizontal flat plate, J. Heat Transf. 117(2) (1995) 309–315. https://doi.org/10.1115/1.2822522.

[25] M.J. Johnson, D.B. Go, Recent advances in electrohydrodynamic pumps operated by ionic winds: a review, Plasma Sources Sci. Technol. 26(10) (2017) 103002. https://doi.org/10.1088/1361-6595/aa88e7.

[26] J. Wang, T. Zhu, Y. Cai, J. Zhang, J. Wang, Review on the recent development of corona wind and its application in heat transfer enhancement, Int. J. Heat Mass Transf. 152 (2020) 119545. https://doi.org/10.1016/j.ijheatmasstransfer.2020.119545.

[27] N.E. Jewell-Larsen, H. Ran, Y. Zhang, M.K. Schwiebert, K.A.H. Tessera, A.V. Mamishev, Electrohydrodynamic cooled laptop, in: Proc. IEEE Semiconductor Thermal Measurement and Management Symp., 2009, pp. 261–266. https://doi.org/10.1109/STHERM.2009.4810773.

[28] C.P. Hsu, N.E. Jewell-Larsen, I.A. Krichtafovitch, A.V. Mamishev, Heat transfer enhancement measurement for microfabricated electrostatic fluid accelerators, J. Microelectromech. Syst. 18(1) (2009) 111–118. https://doi.org/10.1109/JMEMS.2008.2008622.

[29] H.C. Wang, N.E. Jewell-Larsen, A.V. Mamishev, Thermal management of microelectronics with electrostatic fluid accelerators, Appl. Therm. Eng. 51(1) (2013) 190–211. https://doi.org/10.1016/j.applthermaleng.2012.08.068.

[30] A.O. Ongkodjojo, A.R. Abramson, N.C. Tien, Electrohydrodynamic microfabricated ionic wind pumps for thermal management applications, J. Heat Transf. 136(6) (2014) 061703. https://doi.org/10.1115/1.4026807.

[31] D.B. Go, S.V. Garimella, T.S. Fisher, R.K. Mongia, Ionic winds for locally enhanced cooling, J. Appl. Phys. 102(5) (2007) 053302. https://doi.org/10.1063/1.2776164.
[32] D.S. Drew, K.S.J. Pister, Geometric optimization of microfabricated silicon electrodes for corona discharge-based electrohydrodynamic thrusters, Micromachines 8(5) (2017) 141. https://doi.org/10.3390/mi8050141.
[33] G.T. Bellamy, M.J. DiDomizio, M.K. Patel, M.B. McKinnon, Characterization of high-temperature paints for infrared thermography in fire research, Fire Saf. J. 137 (2023) 103775. https://doi.org/10.1016/j.firesaf.2023.103775.
[34] J. Stafford, E. Walsh, V. Egan, Characterizing convective heat transfer using infrared thermography and the heated-thin-foil technique, Meas. Sci. Technol. 20(10) (2009) 105401. https://doi.org/10.1088/0957-0233/20/10/105401.
[35] O. Raghu, J. Philip, Thermal properties of paint coatings on different backings using a scanning photoacoustic technique, Meas. Sci. Technol. 17(11) (2006) 2945. https://doi.org/10.1088/0957-0233/17/11/012.
[36] E.S. Nogueira, J.R.D. Pereira, M.L. Baesso, A.C. Bento, Study of layered and defective amorphous solids by means of thermal wave method, J. Non-Cryst. Solids 318(3) (2003) 314–321. https://doi.org/10.1016/S0022-3093(02)01893-8.
[37] S.W. Churchill, H.H. Chu, Correlating equations for laminar and turbulent free convection from a vertical plate, Int. J. Heat Mass Transf. 18(11) (1975) 1323–1329.
[38] S. Pua, K. Ong, K. Lai, M. Naghavi, Natural and forced convection heat transfer coefficients of various finned heat sinks for miniature electronic systems, Proc. Inst. Mech. Eng. Part J Power Energy 233(2) (2019) 249–261. https://doi.org/10.1177/0957650918784420.
[39] M. Cogollo de Cádiz, A. López Arrabal, A. Díaz Lantada, M.V. Aguirre, Materials degradation in non-thermal plasma generators by corona discharge, Sci. Rep. 11 (2021) 24175. https://doi.org/10.1038/s41598-021-03447-w
[40] N.E. Jewell-Larsen, S.V. Karpov, H. Ran, P. Savalia, K.A. Honer, Investigation of dust in electrohydrodynamic (EHD) systems, in: Proc. IEEE SEMI-THERM, 2010, pp. 249–255. https://doi.org/10.1109/STHERM.2010.5444283
[41] P. Xu, B. Zhang, S. Chen, J. He, Influence of humidity on the characteristics of positive corona discharge in air, Phys. Plasmas 23 (2016) 063511. https://doi.org/10.1063/1.4953890
[42] D.S. Drew, K.S.J. Pister, Takeoff of a flying microrobot with COTS sensor payload using electrohydrodynamic thrust produced by submillimeter corona discharge, in: Solid-State Sensors, Actuators, and Microsystems Workshop, Transducer Research Foundation, 2018, pp. 67–70.
[43] D. Nuvolone, D. Petri, F. Voller, The effects of ozone on human health, Environ. Sci. Pollut. Res. 25(9) (2018) 8074–8088. https://doi.org/10.1007/s11356-017-9239-3.
[44] K. Yanallah, F. Pontiga, A. Fernández-Rueda, A. Castellanos, Experimental investigation and numerical modelling of positive corona discharge: ozone generation, J. Phys. D Appl. Phys. 42(6) (2009) 065202. https://doi.org/10.1088/0022-3727/42/6/065202.
[45] L. Chen, E. Gonze, M. Ondarts, J. Outin, Y. Gonthier, Electrostatic precipitator for fine and ultrafine particle removal from indoor air environments, Sep. Purif. Technol. 247 (2020) 116964. https://doi.org/10.1016/j.seppur.2020.116964.
[46] M.B. Awad, G.S.P. Castle, Ozone generation in an electrostatic precipitator with a heated corona wire, J. Air Pollut. Control Assoc. 25(4) (1975) 369–374.

https://doi.org/10.1080/00022470.1975.10470092.
[47] W.W. Nazaroff, C.J. Weschler, Indoor ozone: concentrations and influencing factors, Indoor Air 32(1) (2022) e12942. https://doi.org/10.1111/ina.12942.